# Bethe-Slater-curve-like behavior and interlayer spin-exchange coupling mechanisms in two-dimensional magnetic bilayers

Cong Wang[1,†], Xieyu Zhou[1,†], Linwei Zhou[1], Yuhao Pan[1], Zhong-Yi Lu[1], Xiangang Wan[2], Xiaoqun Wang[3] and Wei Ji[1,*]

[1]*Beijing Key Laboratory of Optoelectronic Functional Materials & Micro-Nano Devices, Department of Physics, Renmin University of China, Beijing 100872, P.R. China*
[2] *National Laboratory of Solid State Microstructures and School of Physics, Nanjing University, Nanjing 210046, P.R. China*
[3] *Department of Physics and Astronomy, Collaborative Innovation Center of Advanced Microstructures, Shanghai Jiao Tong University, Shanghai 200240, P.R. China*
Corresponding authors: W.J. (email: wji@ruc.edu.cn)

† These authors contributed equally to this work.

Layered magnets have recently received tremendous attention, however, spin-exchange coupling mechanism across their interlayer regions is yet to be revealed. Here, we report a Bethe-Slater-curve (BSC) like behavior in *nine* transition metal dichalcogenide bilayers ($MX_2$, M=V, Cr, Mn; X=S, Se, Te) and established interlayer spin-exchange coupling mechanisms at their van der Waals gaps using first-principle calculations. The BSC-like behavior offers a distance-dependent interlayer anti-ferromagnetic (AFM) to ferromagnetic (FM) transition. This phenomenon is explained with the spin-exchange coupling mechanisms established using bilayer $CrSe_2$ as a prototype in this work. The Se $p_z$ wavefunctions from two adjacent interfacial Se sublayers overlap at the interlayer region. The spin alignment of the region determines interlayer magnetic coupling. At a shorter interlayer distance, Pauli repulsion at the overlapped region dominates and thus favors anti-parallel oriented spins leading to interlayer AFM. For a longer distance, kinetic energy gain of polarized electrons across the bilayer balances the Pauli repulsion and the bilayer thus prefers an interlayer FM state. In light of this, the AFM-FM transition is a result of competition between Pauli and Coulomb repulsion and kinetic energy gain. All these results open a new route to tune interlayer magnetism and the revealed spin-exchange coupling mechanisms are paramount additions to those previously established ones.

Two-dimensional (2D) magnetism has received increasing attention after demonstrations of ferromagnetism (FM) in 2D layers [1-7]. While each magnetic few-layer consists of strong covalently bonded monolayers, their interlayer couplings are governed by much weaker non-covalent, e.g. van der Waals (vdW), interactions at their ``vdW gaps'', which show even more interesting and mysterious magnetic behaviors[1, 8-10]. Bilayer $CrI_3$ is one of the most popular magnetic few-layers, in which local magnetic moments (3.28 $\mu_B$/Cr)[10] form an intra-layer FM order below 45 K[1, 8]. However, its interlayer magnetic coupling is variable between FM and anti-ferromagnetism (AFM) depending on local stacking geometry, highlighting the importance of interlayer magnetic couplings in 2D magnetism [8, 10]. There are numbers of previously well-established spin-exchange coupling mechanisms for classic magnetism, e.g. super-exchange in a linear (AFM)[11] and a perpendicular (FM)[12] configurations, double exchange (FM)[13, 14], direct exchange (AFM/FM)[15, 16] and RKKY interactions (AFM/FM)[17-20]. While these mechanisms were derived mostly on the basis of covalently and/or metallically bonded bulk solids, those for non-covalent interactions at, e.g. vdW gaps, are still lack of knowledge and yet to be unveiled.

Strength of interlayer interactions at vdW gaps was previously thought rather weak but was recently found to be appreciably strong in terms of modifying electronic structures and related physical properties [9, 21-25]. In a significant portion of 2D layers, interlayer wavefunctions, as driven by dispersion attraction, do overlap and hybridize to release Pauli and Coulomb repulsions [9, 21-25]. Charge redistribution induced by this interaction is relatively small at the vdW gap of a $CrI_3$ bilayer[10] in comparison with other bilayers like BP[21, 22], Te[23, 26], $PtSe_2$[24], $PtS_2$[25] and $CrS_2$[9], suggesting a limited overlap of interlayer wavefunctions in the $CrI_3$ bilayer. It is exceptional that such small overlap could even appreciably affect the interlayer magnetism through direct exchange between two interlayer I atoms separated by 4.20 Å [10]. A question then arises that how do strongly overlapped interlayer wavefunctions affect interlayer magnetism and whether there are any generalized spin-exchange coupling mechanism solely for such non-covalent interaction.

Here, we found a Bethe-Slater curve (BSC) [27-30] like behavior at the vdW gaps of *nine* transition-metal di-chalcogenide bilayers (TMDC, $MX_2$, M=V, Cr, Mn; X=S, Se, Te) using density functional theory (DFT) and unveiled two spin-exchange coupling mechanisms, together with a modified Hubbard model, to understand such behavior. In particular, each of these bilayers prefers interlayer AFM at shorter interlayer distances and favors FM at elongated distances. It was found to determine the AFM-FM transition that a competition between Pauli (Coulomb) repulsion at the interlayer region and kinetic energy gain across the entire bilayer. This behavior and the explanation are general for different chalcogen and/or transition metal atoms.

We use $CrSe_2$ as a prototypic TMDC in this work. Its monolayer takes a hexagonal 1T structure with the P-3M1 space group (Fig. 1a), which is an analogue of the strongly interlayer-coupled $CrS_2$ layers[9], but offers better synthetic feasibility[31, 32]. Computational details can be seen in the method part of supplementary material[33] (see, also, references [21-23, 25, 34-49] therein). Its FM order is meta-stable with an optimized lattice constant of *a*=3.42 Å and magnetic moment of *M*=3.09 $\mu_B$/Cr (Fig. S1 and Table S1[33]). The groundstate is, however, striped AFM (sAFM-ABAB, Fig. S2c[33]) in an 1×$\sqrt{3}$ orthorhombic lattice with slightly expanded *a*=3.50 Å, largely shrank *b*=5.63 Å and nearly unchanged *M*=3.02 $\mu_B$/Cr, see Table S1[33] for details. Competition of in-plane spin-exchange parameters $J_1$ = -2.32 meV and $J_2$ = -0.91 meV leads to the sAFM-ABAB groundstate. In a $CrSe_2$ bilayer, the AA stacking (Fig. 1b and 1c) is over 13.9 meV/Cr more stable than the AB stacking (Fig. S3 and Table S2[33]) and was chosen for further discussion. It undergoes an intralayer sAFM-to-FM transition in the bilayer ($J_1$ = 7.70 meV and $J_2$ = 1.60 meV), ascribed to the strong interlayer wavefunction overlap induced Cr $e_g$-to-$t_{2g}$ charge transfer and its resulting intralayer double-exchange mechanism [9]. This strong coupling is confirmed by a ~0.2 Å reduced monolayer thickness, a ~0.1 $\mu_B$/Cr enhanced magnetic moment and a large interlayer binding energy $E_b$=-0.32 eV/formula unit (f.u.) (see Table S1[33]). While bilayer $CrS_2$ shows interlayer FM, interlayer AFM configuration in bilayer $CrSe_2$ is 1.16 meV/Cr more favored than the interlayer FM. This unexpected interlayer AFM is determined by parameters $J_3$ - $J_5$ (Fig. 1b-c and Table S1[33]). Although parameter $J_3$ has the nearest interlayer Cr-Cr distance of 5.67 Å, the farthest (8.31 Å) parameter $J_5$=-1.25 eV, however, yields the strongest AFM coupling strength while $J_3$=0.95 eV and $J_4$=0.90 eV show weaker FM couplings. All these results suggest the CrSe$_2$ bilayer deserves a closer examination.

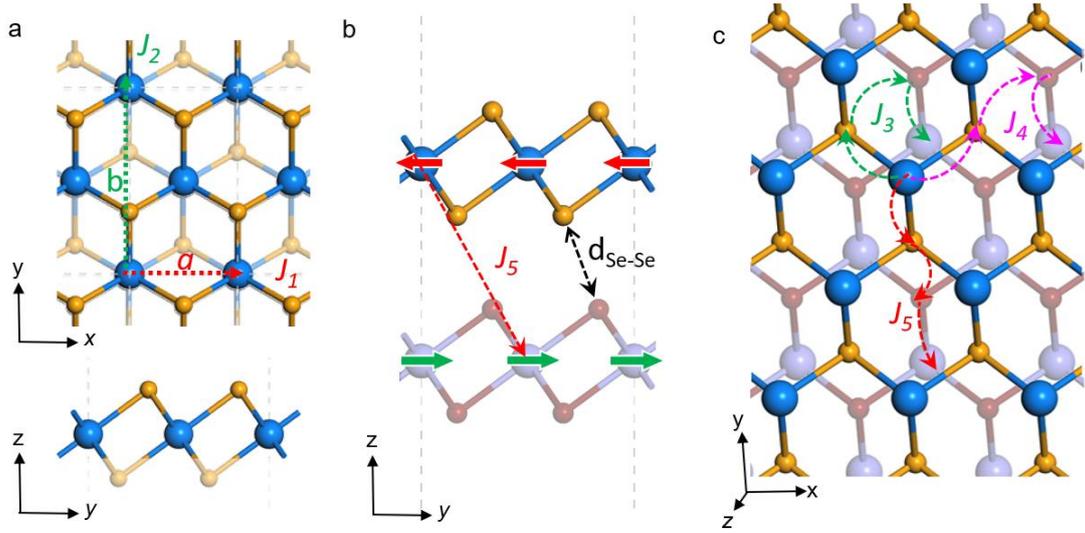

**Figure 1. Atomic structures of CrSe$_2$ mono- and bi-layers.** (a) Top and side views of a CrSe$_2$ monolayer. Dodger-blue, solid and semitransparent orange balls represent Cr, top and bottom Se atoms, respectively. Red and green dashed arrows denote intralayer spin-exchange parameters $J_1$ and $J_2$ between Cr sites, corresponding to lattice constants $a$ and $b$. (b) Side-view of an AA-stacked CrSe$_2$ bilayer. Green and red arrows represent magnetization directions of Cr atoms. Violet and dark red balls represent Cr and Se atoms in the bottom layer, respectively. The red dashed arrow indicates exchange pathway $J_5$. The black arrow represents the interlayer Se-Se distance d$_{Se-Se}$. (c) Top oblique view of bilayer CrSe$_2$ explicitly shows three exchange pathways of $J_3$, $J_4$ and $J_5$ with green, pink and red dashed arrows.

The interlayer magnetism of the CrSe$_2$ bilayer strongly depends on vertical sliding, while laterally sliding in a CrI$_3$ bilayer allows tuning its interlayer magnetism between AFM and FM[10, 50, 51]. Figure 2a plots the interlayer AFM-FM energetic difference as a function of the interlayer Se-Se distance ($d_{Se-Se}$, Fig. 2a inset), which shows a Bethe-Slater curve (BSC)[27-30] like behavior at the vdW gap. The BSC was used to explain different magnetic orderings of metals, e.g. Cr[52], Ni[53], Fe[54]. All considered functionals and vdW correction methods [38-40, 43-48] exceptionally show that interlayer AFM and FM are favored at shorter and longer distances, respectively, although different functional slightly affect the AFM-FM transition distance ($d_T$) (Table S5[33]). Particularly, PBE predicts $d_T$=3.45 Å while HSE06 defers it by 0.05 Å. Spin-orbit coupling (SOC) reduces the energy difference by 1-2 meV/Cr but keeps $d_T$ nearly unchanged (Fig. 4a and Table S5[33]). Effects of magnetic anisotropy energies (MAEs) were also considered that it is one order of magnitude smaller than the AFM-FM energy differences, which unlikely influence the ground state of a certain bilayer (Table S6[33]).

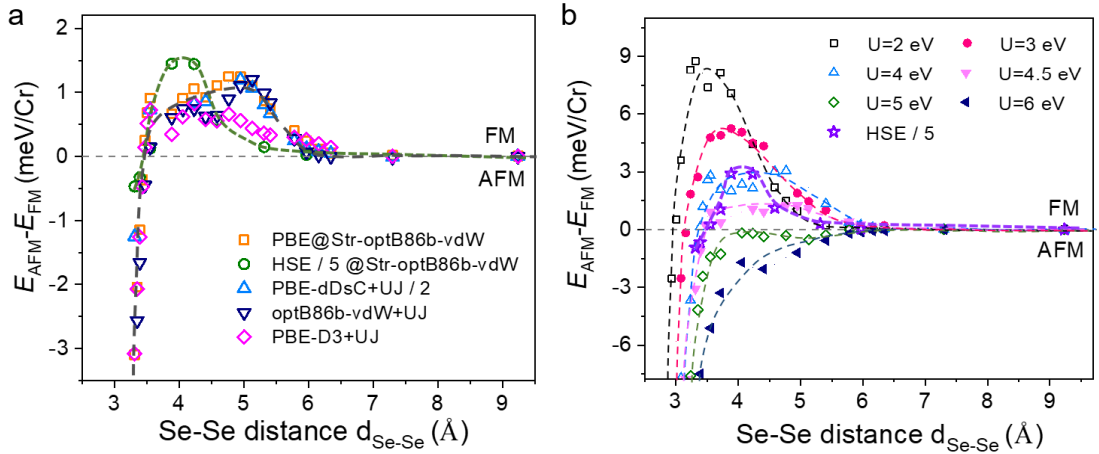

**Figure 2. Bethe-Slater-curve-like behavior in bilayer CrSe$_2$.** (a) Total energy differences between the interlayer AFM/FM configurations as a function of interlayer Se-Se distance $d_{Se-Se}$ (marked with black arrow in Fig. 1b) for the CrSe$_2$ bilayer calculated using different functionals. Here, PBE and HSE06 energies were compared based on optB86b-vdW relaxed structures. Energy differences for the PBE-dDsC+UJ and the HSE06 results are plotted with scaling factors of 1/2 and 1/5, respectively. The inset shows a perspective view of the AA-stacked CrSe$_2$ bilayer and the definition of the Se-Se distance. (b) FM-AFM energy differences evolution with different on-site $U$ values with a constant $J$ value of 0.6 eV. The HSE06 results were plotted for reference.

Figure 2b plotted the $U$ dependence of distance-dependent FM-AFM energy differences and Table S7[33] summarized the dependence of $d_T$ and magnetic moments. A smaller $U$-value of 2.0~4.0 eV shortens $d_T$, however, a larger $U$-value of 5.0~6.0 eV entirely eliminates the FM region and enlarges the energy difference by 2.01 meV/Cr and magnetic moment of 0.05 $\mu_B$/Cr at the equilibrium distance ($d_E$). Although $d_T$ depends on the $U$-value, the transition behavior is well reproduced by the HSE06 functional (green circles in Fig. 2a and purple stars in Fig. 2b), which is usually believed to better predict magnetic properties than DFT+U methods do. Here, the linear-response method [49] derived $U$=4.5 eV gives the upper limit of $U$. We are thus confident to claim the BSE-like behavior in the CrSe$_2$ bilayer. This $U$-value dependent $d_T$ suggests the transition is, most likely, relevant with Coulomb and Pauli interactions and electron kinetic energy because of their $U$-dependence in DFT calculations.

While the in-plane orbitals of Se primarily determine the intralayer magnetism, the out-plane Se 4$p_z$ orbitals may play a paramount role in interlayer magnetism. We thus focus on the role of Se 4$p_z$ orbitals. In the CrSe$_2$ bilayer, two interfacial Se 4$p_z$ orbitals overlap and hybridize into one bonding and one anti-bonding states (Fig. S4[33]). We mapped the wavefunction norm of this bonding state on the atomic structure in Fig. 3a where an explicit overlapped region (OR) is evidenced at the interlayer area, which could be effectively considered as an area accumulating appreciable shared charge from the two interfacial Se sublayers.

In the interlayer FM configuration (Fig. 3b-3f), interfacial Se (Se_it, Se_ib) 4$p_z$ orbitals significantly overlap (denoted by curved short-dashed lines in Fig. 3b and 3c), leading to charge transfers of the spin-1 (up) component from interfacial Se 4$p_z$ to Cr $t_{2g}$ (indicated by curved red solid arrows in Fig. 3b). As a result, the third Cr_bot(top) $t_{2g}$ orbital becomes partially occupied (Fig. 3c) and the averaged local magnetic moment of Cr enlarges by 0.05 $\mu_B$, see Table S1[33], which lead to the intralayer sAFM to FM transition from monolayer to bilayer through Cr-Cr double-exchange, similar to the CrS$_2$ case[9]. Those charge transfers are supported by differential charge density (DCD) plots (Fig. S5[33]). In addition, the transferred spin-1 (spin-up for simplification hereinafter) charge of Se $p_z$ to Cr leaves the spin-2 (spin-down hereinafter) component predominated at the OR, as shown in Fig. 3e and 3f and illustrated in Fig. 3c.

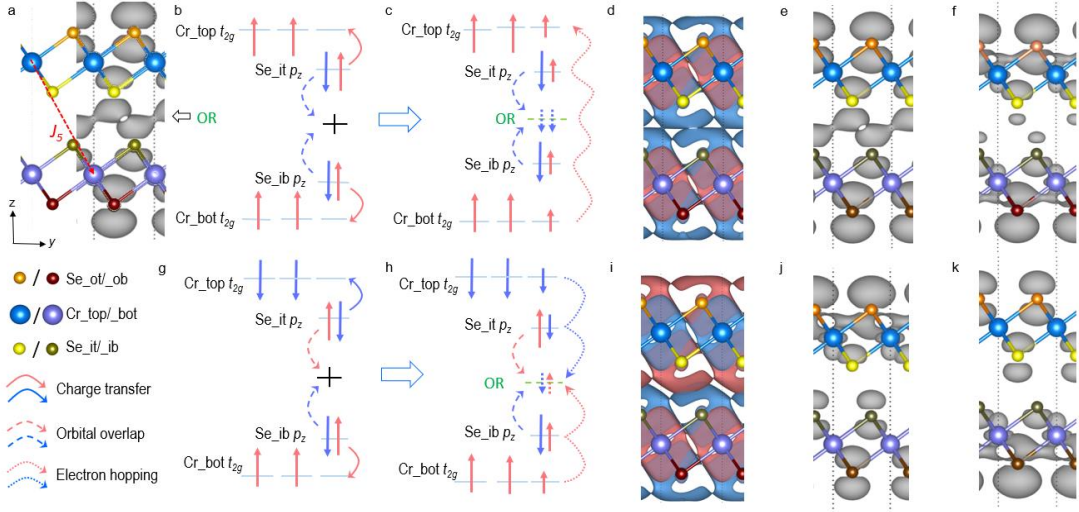

**Figure 3. Spin-exchange coupling mechanisms of the CrSe$_2$ bilayer.** (a) Side view of bilayer CrSe$_2$ with indicative of spin-exchange coupling $J_5$ and mapped with the wavefunction norms of the Se 4$p_z$ bonding state where an overlapped region (OR) was identified. Atoms at different positions are marked with different colors. (b,g) are schematics of the interlayer charge transfer and wavefunction overlaps and (c,h) illustrate for spin-exchange mechanisms of the interlayer FM (b,c) and AFM (g,h) configurations, respectively. Spin-up and -down electrons are represented by red straight-up and blue straight-down arrows, the length of which qualitatively represents the amount of spin-polarized electrons. Red (spin-up) and blue (spin-down) curved arrows indicate the charge transfer (solid), wavefunction overlap (dashed) and electron hopping (dotted), respectively. Spin densities of the both configurations were plotted in (d) and (i), respectively, with an isosurface value of 0.0004 $e$/Bohr$^3$. Red and blue contours denote spin-up and -down, respectively. The maximum value of the spin density is 0.6 e/Bohr$^3$, locating at the Cr site. (e,f) and (j,k) visualized the wavefunction norms of the spin-down (e,j) and –up (f,k) components of the interfacial bonding states for both interlayer FM (e,f) and AFM (j,k), respectively. An isosurface value of 0.0015 e/Bohr$^3$ was used.

We then focus on its interlayer spin-exchange coupling mechanism for $J_5$, as depicted in Fig. 3c. The third $t_{2g}$ orbitals of both Cr_top and Cr_bot are spin-up polarized and partially occupied. The spin-up electrons of Cr_bot $t_{2g}$, for example, could hop into the Se_ib 4$p_z$ orbital and then reach the OR upon excitation because they are spin-down polarized and still allow spin-up electrons to occupy. Given the same reason, the electron could go further from the OR to Cr_top $t_{2g}$ through Se_it 4$p_z$, as denoted by the wave-like red dotted arrow; this substantially lowers the kinetic energy of spin-up electrons across the bilayer. This process is similar to double-exchange but is mediated by multiple sites. We thus termed it ``multi-intermediate double-exchange". However, we should notice that the OR are effectively filled with two spin-down fractional electrons, which violates the Pauli exclusion law if OR is a real atomic site, giving rise to an appreciable Pauli repulsive interaction ($P$). We used a modified interlayer Hubbard model to describe it as follows,

$$H = - \sum_{i,j,\sigma=\downarrow,\uparrow} t_{ij}(C_{i\sigma}^+ C_{j\sigma} + h.c.) + \sum_i U_i n_{i\uparrow} n_{i\downarrow} + \sum_{\sigma=\downarrow,\uparrow} P_\sigma n_{or\_t,\sigma} n_{or\_d,\sigma}$$

where the first and second terms are hopping and on-site Coulomb contributions of the conventional Hubbard model and the third term represents Pauli repulsion at the OR (more strictly, between the two $p_z$ orbitals of Se_it and Se-ib). Here, $i$ and $j$ span all four atomic sites, i.e. Cr_top(bot) and Se_it(b), while the Pauli term is for the OR solely and $n_{es\_t(or\_b),\sigma}$ represents spin-dependent occupation at the OR contributed from Se_it (ib), see Supplementary Note 1[33] (see, also, references [55, 56] therein) for details. In other words, the third Pauli term accounts for the Pauli repulsive energy between the overlapped $p_z$ wavefucntions of Se_ib and Se_it. Because the CrSe$_2$ bilayer is metallic for the interlayer direction (Fig. S7[33]), it was usually believed $t > U$. If Pauli repulsion $P$ could be further surmounted by hopping $t$, the interlayer FM configuration is still favored, otherwise, an interlayer AFM configuration is suggested.

The AFM bilayer also has the stacking induced charge transfer and the interfacial wavefunction overlap, but both spin-components are involved (Fig. 3g). Thus, the overlapped wavefunction at the OR is composed of both spin components (Fig. 3h-3k). Charge transfers occurred in the bottom and top CrSe$_2$ layers (denoted by solid arrows in Fig. 3g) give rise to more strongly polarized Cr_bot and Cr_top $t_{2g}$ orbitals, anti-parallel polarized Se_it and Se_ib 4$p_z$ orbitals and the non-polarized OR, respectively, as illustrated in Fig. 3h and depicted by spin-density in Fig. 3i and spin-dependent wavefunction norms in Fig. 3j and 3k. Figure 3h shows two major differences for the AFM spin-exchange coupling mechanism from the FM one. The spin non-polarized OR indicates that Pauli repulsion $P$ is largely eliminated, which substantially lowers the total energy of the bilayer. However, such configuration shortens the range that a spin-polarized electron can move across the bilayer. We use a spin-down electron of Cr_top $t_{2g}$ (in blue) as an example. It could hop into the Se_it $4p_z$ site and then to the OR, similar to the FM case, but further hopping to the Se_ib $4p_z$ site is forbidden because its spin-down component is fully occupied; this appreciably lifts up the kinetic energy. Again, if the lowered potential energy by eliminating Pauli repulsion overcomes the risen kinetic energy, AFM is thus preferred as the interlayer magnetic groundstate.

In short, competition between the interlayer hopping ($t$) across the bilayer and the Pauli ($P$) and Coulomb ($U$) repulsions at the OR determines the interlayer magnetism of CrSe$_2$, resulting in the BSC-like behavior (Fig. 2). This phenomenological picture is supported by the calculations of the interlayer-distance dependent exchange-splitting (Coulomb), bandwidth (hopping) and spin-polarized electron-density (Pauli) of the interfacial Se $p_z$ orbitals (Fig. S6[33]). At shorter distances, the FM state shows enlarged exchange-splitting of the $p_z$ orbital, increased density of the same spin component at the OR and reduced bandwidth of the interlayer bonding state. All these results prefer the AFM state. Nevertheless, it is prone to favor FM with increasing interlayer distance. An enlarged $U$-value localizes electrons and thus reduces the role of kinetic energy but enhances that of Pauli repulsion, more preferring interlayer AFM, consistent with our results shown in Fig. 2b. At $d_T$ of bilayer CrSe$_2$, the more predominant Pauli repulsion finds the interlayer AFM groundstate with 0.05 Å shorter interlayer distance and 0.01 eV larger interlayer binding energy.

We extended our discussion to CrS$_2$ and CrTe$_2$ bilayers. Figure 4a plots the energy-distance relations of CrX$_2$ (X=S, Se, Te) revealed using HSE06(-SOC), each of which follows the expected BSC-like behavior with different $d_T$ values. The $d_T$ of 2.74 Å for CrS$_2$ is 0.46 Å shorter than its $d_E$ of 3.20 Å, at which it nearly yields the largest energy difference, consistent with the strong FM in the literature[9]. Although $d_T$ of CrTe$_2$ depends on the functional or U value adapted or whether SOC was considered (Fig. 4b and Table S8[33]), the smallest predicted $d_T$ is 4.22 Å, much larger than its $d_E$ of 3.62 Å, implying the interlayer AFM in a broad spectrum of distance. Competition between Pauli and Coulomb repulsions results in this element-dependent transition-equilibrium distances relation, the details of which were discussed in Figs. S7, S8 and Table S3, S8[33] (see, also, references [57-61] therein). We further generalized the BSC-like behavior to $VX_2$[6] and $MnX_2$[4, 62] bilayers, all of which follow our expectation (Fig. 4c and Fig. S8[33]). The BSC-like behavior maintains even a monoclinic 1T" phase[61], as a special case, was considered for $VTe_2$ (Fig. S9 and Table S4[33]), suggesting its robustness.

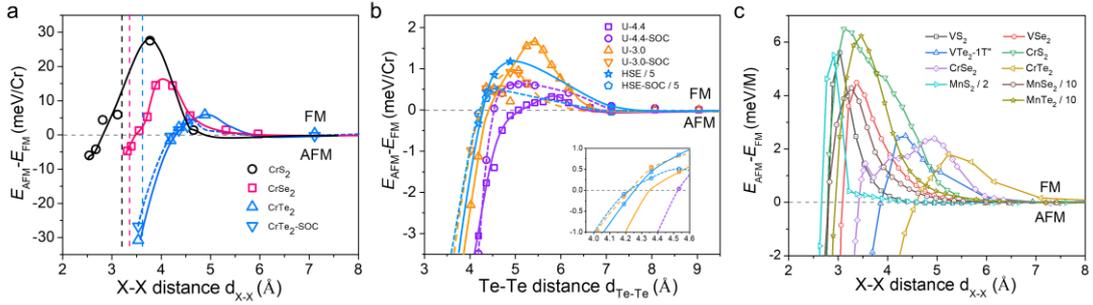

**Figure 4. Generalization of Bethe-Slater-curve-like behavior.** (a) Distance dependent FM-AFM energy differences revealed with HSE06 for $CrS_2$ (black), $CrSe_2$ (red), and $CrTe_2$ (blue) bilayers. Equilibrium positions of $CrS_2$, $CrSe_2$ and $CrTe_2$ are marked with black, red and blue dashed vertical lines. (b) Comparison of the BSC-like curves revealed with different on-site U values with the HSE06 ones for bilayer $CrTe_2$. All SOC results were shown in dashed lines while results without SOC are represented by solid lines. (c) AFM-FM energy differences for $MX_2$ (M=V, Cr, Mn; X=S Se, Te) bilayers calculated with PBE-dDsC+UJ (See Table S9[33] for structural details).

In summary, we proposed a modified Hubbard model to explain the interlayer FM and AFM ground-states for nine $MX_2$ M=V, Cr, Mn; X=S, Se and Te) bilayers at different distances. Each of these bilayers contains an interlayer wavefunction overlapped region. Competition between Pauli and Coulomb repulsions at this region and kinetic energy gain across the bilayer leads to the BSC-like behaviors, i.e. distance-dependent AFM-FM transition, in all considered $MX_2$ bilayers. Differently from the BSC behavior in 3D magnetic materials, only a 0.05 Å increasement of interlayer distance, with an energy cost of tens of meV, could induces the AFM to FM transition, illustrating the feasibility of tuning magnetism by changing interlayer distance. While

``Multi-intermediate double-exchange'' was termed for the interlayer FM coupling, for the interlayer AFM coupling, we could regard the X_it-OR-X_ib group as a super-orbital and M_top and M_bot atoms couples through this ``super-orbital mediated super-exchange'' mechanism. We can infer that these two interlayer magnetic coupling mechanisms also work for other strongly electronic coupled 2D magnetic layers that S, Se or Te atoms sit between their vdW gaps, like $Fe_3GeTe_2$ and $MnBi_2Te_4$. In addition, these mechanisms also suggest a giant magneto-elastic effect in 2D layers where a small interlayer vertical displacement changes the total magnetism.

## Acknowledgements

We gratefully acknowledge financial support from the Ministry of Science and Technology (MOST) of China (Grant No. 2018YFE0202700), the National Natural Science Foundation of China (Grant No. 11622437, No. 61674171 and No. 11974422), the Strategic Priority Research Program of Chinese Academy of Sciences (Grant No. XDB30000000), the Fundamental Research Funds for the Central Universities, China, and the Research Funds of Renmin University of China [Grants Nos. 16XNLQ01 and 19XNQ025 (W.J.) and No. 19XNH065 (X.Z.)]. Cong Wang was supported by the Outstanding Innovative Talents Cultivation Funded Programs 2017 of Renmin University of China. Calculations were performed at the Physics Lab of High-Performance Computing of Renmin University of China, Shanghai Supercomputer Center. During the review process of this manuscript, we were aware that an extended super-exchange mechanism was proposed for address the magnetism in bulk Ca3LiOsO6[63], which is essentially similar to our proposed ``super-orbital mediated super-exchange''. We thank Dr. Yanfeng Guo for bringing this reference to us.

# Supplemental Information

# Bethe-Slater-curve-like behavior in double super-exchange governed two-dimensional magnetic bilayers

Cong Wang[1,†], Xieyu Zhou[1,†], Linwei Zhou[1], Yuhao Pan[1], Zhong-Yi Lu1, Xiangang Wan[2], Xiaoqun Wang[3] and Wei Ji[1,*]

[1]*Beijing Key Laboratory of Optoelectronic Functional Materials & Micro-Nano Devices, Department of Physics, Renmin University of China, Beijing 100872, P.R. China*
[2]*National Laboratory of Solid State Microstructures and School of Physics, Nanjing University, Nanjing 210046, P.R. China*
[3] *Department of Physics and Astronomy, Collaborative Innovation Center of Advanced Microstructures, Shanghai Jiao Tong University, Shanghai 200240, P.R. China*
Corresponding authors: W.J. (email: wji@ruc.edu.cn)
† These authors contributed equally to this work.

# Computational methods

Our density functional theory calculations were performed using the generalized gradient approximation and the projector augmented wave method [1, 2] as implemented in the Vienna *ab-initio* simulation package (VASP) [3, 4]. The uniform Monkhorst-Pack k mesh of 15×15×1 was adopted for integration over Brillouin zone. An orthorhombic $1\times\sqrt{3}$ supercell was used to show the stripe antiferromagnetic order in monolayer $CrSe_2$, with the k mesh of 16×10×1 was adopted respectively. A plane-wave cutoff energy of 700 eV was used for structural relaxation and electronic structures. A sufficiently larger distance of $c > 20$ Å along out-of-plane direction was adopted to eliminate interaction between each layer. Dispersion correction was made at the vdW-DF level [5-7], with the optB86b functional for the exchange potential, which was proved to be accurate in describing the structural properties of layered materials [8-13] and was adopted for structure related calculations. All atoms were allowed to relax until the residual force per atom was less than 0.01 eV/ Å. The results of geometric relaxation were double checked with PBE-dDsC functional [14, 15]. Little difference was found between structures calculated using these two functional (**Table S10**).

For energy comparisons among different magnetic configurations, we used the PBE [16] and HSE06 functionals [17, 18], with consideration of spin-orbit coupling (SOC), based on the vdW-DF optimized atomic structures. These results were also double checked with the PBE-dDsC functional [14, 15] and the PBE-D3 [19]. A plane-wave cutoff energy of 600 eV was used for energy calculations. On-site Coulomb interaction energies to the Cr $d$ orbitals were self-consistently calculated with a linear-response method [20]. This calculation reveals $U = 4.5$ eV and $J = 0.6$ eV for $CrSe_2$ (see **Table S11** for details of other bilayers), which were used in our calculations except for HSE06. A Heisenberg model was used to model the magnetism of the bilayer,

$$H = H_0 - \sum_{k=1}^{5} J_k \sum_{i,j} \vec{S_i} \cdot \vec{S_j}$$

which includes two in-plane spin-exchange interactions $J_1$ and $J_2$ (**Fig. 1a**), three interlayer ones, $J_3$, $J_4$ and $J_5$ (**Fig. 1c**). Expressions of magnetic interaction energies in different magnetic orders are illustrated in **Fig. S2**. All energy comparisons were made with the relaxed interlayer FM structure, unless otherwise noted, to avoid energy variance caused by structural difference. The BSC-like behavior and the interlayer AFM-FM transition were also double checked with interlayer AFM configuration, which yields qualitatively the same results. We do not change in-plane geometric structures when comparing the interlayer AFM-FM energy differences ($E_{AFM}$-$E_{FM}$) as a function of interlayer X-X distance (X= S, Se, Te). All methods adopted, i.e. standard

Perdew-Burke-Ernzerhof (PBE) [16], optB86b-vdW [5-7], density-dependent-dispersion-corrected PBE (PBE-dDsC)[14, 15], Grimme zero-damping-D3-form-corrected PBE (PBE-D3)[19] and Heyd-Scuseria-Ernzerhof (HSE06)[17, 18] functionals, exceptionally show that interlayer AFM and FM are favored at shorter and longer distances, respectively, although different functional slightly affect the AFM-FM transition distance ($d_T$) (Figure 2a and Table S5).

## Supplementary Note1

We used a modified interlayer Hubbard model to describe the spin-exchange coupling mechanism as follows,

$$H = -\sum_{i,j,\sigma=\downarrow,\uparrow} t_{ij}(C_{i\sigma}^+ C_{j\sigma} + h.c.) + \sum_i U_i n_{i\uparrow} n_{i\downarrow} + \sum_{\sigma=\downarrow,\uparrow} P_\sigma n_{or\_t,\sigma} n_{or\_d,\sigma}$$

where the first and second terms are hopping and on-site Coulomb contributions of the conventional Hubbard model and the third term represents Pauli repulsion. Operator $C_{i\sigma}^+$ creates an electron with spin $\sigma$ on site $i$ and operator $C_{i\sigma}$ annihilates an electron with spin $\sigma$ on site $i$. Here, $i$ and $j$ span all four real atomic sites, i.e. Cr_top(bot), Se_it(b) and OR, while the Pauli term is for the overlapped region (OR) solely. While $t_{ij}$ denotes the hopping between sites $i$ and $j$ for electron of the same spin, $U_i$ describes Coulomb interaction on site $i$ for different spins. Operator $n_{i\sigma}$ is the number operator, counting the occupation on site $i$ for a given spin $\sigma$. $P_\sigma$ gives the Pauli repulsion of electrons of the same spin $\sigma$ at the OR, which is element dependent. Strictly speaking, those electrons are from overlapped orbitals of adjacent sites, i.e. Se_it and Se_ib. Operator $n_{or\_t(or\_b),\sigma}$ thus represents spin-dependent occupation on the OR contributed from Se_it (ib).

The idea of the Pauli term in the modified Hubbard model is in analogue to the Pauli repulsive interaction employed in q-plus non-contact Atomic Force Microscopy (nc-AFM) imaging[21, 22]. In nc-AFM imaging, a high electron density apex atom, as the tip, is driven at a certain frequency to probe the electron density of the sample upon approaching the tip to the sample. Strong, but unpleased, wavefunction overlap between the tip and the sample gives rise to an increased energy and changes the vibrational frequency of the tip. In 2D layers, the wavefunction overlap occurs at the interlayer region and is driven by the dispersion attraction and other external fields.

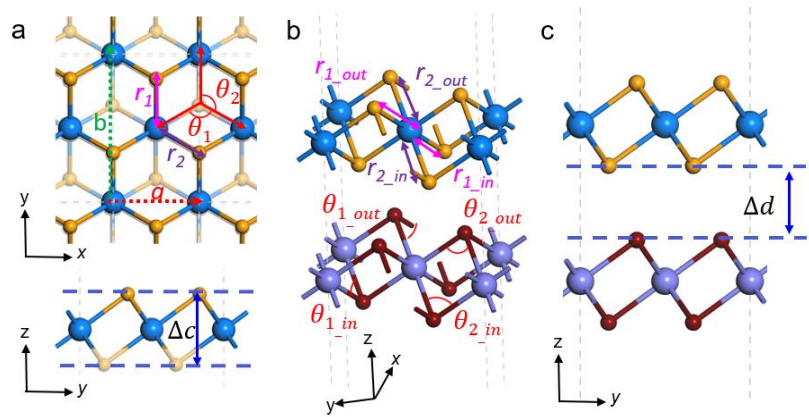

**Figure S1. Schematic models of the CrSe$_2$ mono- and AA stacked bi-layer.** Lattice constants *a* and *b* are marked with red and green dashed arrows in Fig. S1a, respectively. Δ*c* standards for the thickness of each layer (marked in the lower panel of Fig. S1a) while Δ*d* represents the vertical distance between two interlayer Se atoms (marked in Fig. S1c). The detailed values of lattice constant, bond length and angle marked here are listed in Table S1.

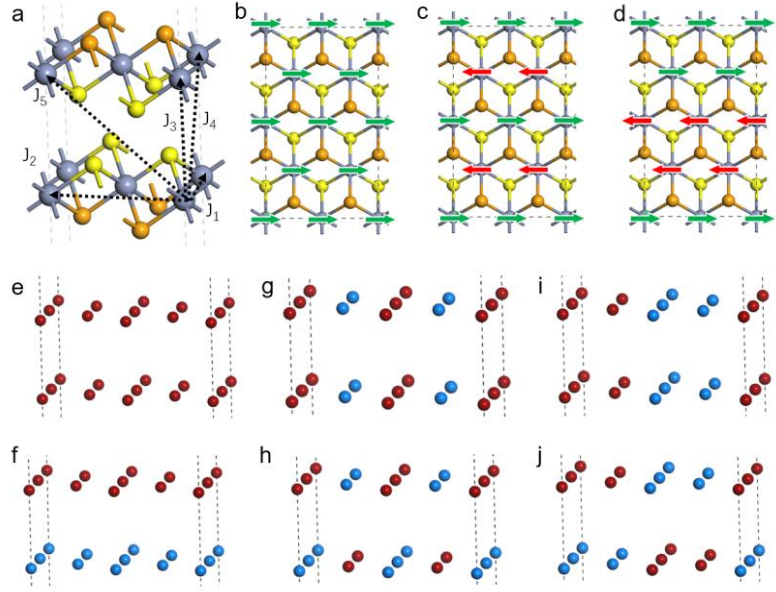

**Figure S2. Spin-exchange parameters of mono-/bi-layer CrSe$_2$.** (a) Perspective view of AA-stacked bilayer CrSe$_2$. Intralayer/interlayer exchange parameters $J_1$-$J_5$ are marked with black dashed arrows bridging two Cr sites. Light-slate-gray, orange and yellow balls represent Cr atoms, outermost and interfacial Se atoms, respectively. (b-d) Top views of schematics showing intralayer magnetic orders, including FM (b), sAFM-ABAB (c) and sAFM-AABB (d) in monolayer/bilayer CrSe$_2$, respectively. (e-j) Schematics of six magnetic orders used for deriving spin-exchange parameters of bilayer CrSe$_2$. Green and red balls represent two anti-parallel orientations of magnetic moments on Cr atoms, respectively. Magnetic energies of these magnetic configurations in each magnetic unit cell read as follow:

$$E_e = E_0 - \frac{N^2}{4} \times \frac{1}{2}(6J_1 + 6J_2 + J_3 + 6J_4 + 6J_5)$$

$$E_f = E_0 - \frac{N^2}{4} \times \frac{1}{2}(6J_1 + 6J_2 - J_3 - 6J_4 - 6J_5)$$

$$E_g = E_0 - \frac{N^2}{4} \times \frac{1}{2}(-2J_1 + 6J_2 + J_3 - 2J_4 - 2J_5)$$

$$E_h = E_0 - \frac{N^2}{4} \times \frac{1}{2}(-2J_1 + 6J_2 - J_3 + 2J_4 + 2J_5)$$

$$E_i = E_0 - \frac{N^2}{4} \times \frac{1}{2}(2J_1 - 2J_2 + J_3 + 2J_4 - 2J_5)$$

$$E_j = E_0 - \frac{N^2}{4} \times \frac{1}{2}(2J_1 - 2J_2 - J_3 - 2J_4 + 2J_5)$$

Here, N represents the number of unpaired spins on each Cr atom, which is chosen as 3 in our calculations.

**Table S1.** Geometric and magnetic details of the mono- and bi-layer $CrSe_2$ calculated with optB86b-vdW+UJ. Here, sAFM and FM stand for striped anti-ferromagnetic and ferromagnetic, respectively. $N_{Layer}$ shows the number of layers, $\Delta E$ is the energy difference from the ground state. Term $E_b$ shows the binding interlayer binding energy. The lattice constants, bond lengths and angles marked in Fig. S1 were listed here. $Se_{in}$ and $Se_{surf}$ represent the interfacial and outermost surface Se atoms, respectively. Magnetic moments were projected onto each atom and the spin-exchange coupling parameters $J_1$-$J_5$ were derived according to the detailed formulism shown in Fig. S2. Here, a negative value means AFM coupling. All energy and magnetic moment related values were calculated using PBE+UJ.

| $N_{Layer}$ | | 1L | | 2L | |
|---|---|---|---|---|---|
| Intra/Interlayer Mag. Config | | FM/- | sAFM/- | FM/FM | FM/AFM |
| $\Delta E$ (meV/Cr) | | 9.29 | 0.00 | 1.16 | 0.00 |
| $E_b$ (eV) | | - | - | -0.31 | -0.32 |
| Str.constants (Å) | a | 3.42 | 3.50 | 3.53 | 3.54 |
| | b | 5.93 | 5.63 | 6.11 | 6.15 |
| | $\Delta c$ | 3.19 | 3.25 | 3.01 | 2.99 |
| | $\Delta d$ | - | - | 2.73 | 2.67 |
| Bond_length (Å) | $r_{1\_out}$ | 2.54 | 2.52 | 2.54 | 2.54 |
| | $r_{1\_in}$ | - | - | 2.53 | 2.53 |
| | $r_{2\_out}$ | 2.54 | 2.55 | 2.54 | 2.54 |
| | $r_{2\_in}$ | - | - | 2.53 | 2.53 |
| Angle (°) | $\theta_{1\_out}$ | 85.03 | 86.84 | 88.32 | 88.58 |
| | $\theta_{1\_in}$ | - | - | 88.56 | 88.74 |
| | $\theta_{2\_out}$ | 85.03 | 81.69 | 88.32 | 88.58 |
| | $\theta_{2\_in}$ | - | - | 88.56 | 88.74 |
| MAG.Mom ($\mu_B$) | Cr | 3.09 | 3.02 | 3.11 | 3.14 |
| | $Se_{out}$ | -0.21 | -0.04 | -0.20 | -0.21 |
| | $Se_{surf}$ | - | - | -0.22 | -0.20 |
| Exchange parameter (meV/Cr) | $J_1$ | -2.32 | | 7.70 | |
| | $J_2$ | -0.91 | | 1.60 | |
| | $J_3$ | - | | 0.95 | |
| | $J_4$ | - | | 0.90 | |
| | $J_5$ | - | | -1.25 | |

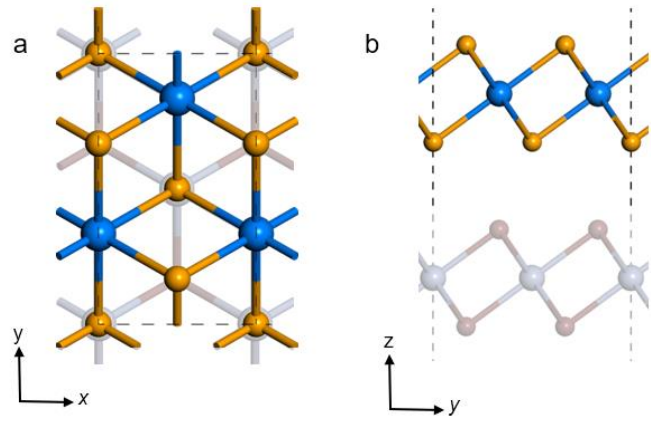

**Figure S3. Schematic models of the CrSe$_2$ bilayer in AB stacking.** Stacking orders AA (Fig. 1) and AB (shown here) are two usually considered stacking orders in TMDs. In terms of bilayer CrSe$_2$, AA stacking configurations were found to be more stable than the corresponding AB configurations. Detailed values of relative energies are available in Table S2.

**Table S2.** Relative total energies and geometry details of bilayer CrSe$_2$ with the AA or AB stacking order in different magnetic configurations shown in Figure S1. Here, FM stands for ferromagnetic [see. Fig. S2(b)], sAFM-ABAB and -AABB refer to two striped anti-ferromagnetic configurations shown in Fig. S2(c) and S2(d), respectively. Term ``Mag. Confg.'' Is the abbreviation of magnetic configuration. Two functionals, i.e. optB86b-vdW and PBE with consideration of SOC and on-site Coulomb interactions (PBE+U-SOC), were used to compare relative energies. For each functional, we set the total energy of the FM-AFM configuration as the reference zero and $\triangle$E represents the different between this configuration of a certain one. Structural parameters, with details marked in Fig. S1, are also available for each configuration.

| 2L Stacking | Mag. Confg. | | $\triangle$E(meV/Cr) | | a(Å) | b(Å) | $\triangle$c(Å) | $\triangle$d(Å) |
|---|---|---|---|---|---|---|---|---|
| | intralayer | interlayer | optB86b-vdW+UJ | PBE+UJ-SOC | | | | |
| AA | FM | FM | 3.16 | 1.16 | 3.53 | 6.11 | 3.01 | 2.73 |
| | FM | AFM | 0.00 | 0.00 | 3.54 | 6.15 | 2.99 | 2.67 |
| | sAFM-ABAB | FM | 17.20 | 30.56 | 3.54 | 5.93 | 3.09 | 2.75 |
| | sAFM-ABAB | AFM | 16.86 | 32.22 | 3.55 | 5.90 | 3.09 | 2.76 |
| | sAFM-AABB | FM | 13.71 | 19.76 | 3.54 | 5.98 | 3.07 | 2.78 |
| | sAFM-AABB | AFM | 19.98 | 25.02 | 3.52 | 6.04 | 3.05 | 2.79 |
| AB | FM | FM | 33.04 | 23.84 | 3.49 | 6.08 | 3.06 | 2.94 |
| | FM | AFM | 29.85 | 21.26 | 3.51 | 6.07 | 3.06 | 2.91 |
| | sAFM | FM | 14.61 | 49.39 | 3.51 | 5.63 | 3.24 | 3.31 |
| | sAFM | AFM | 13.91 | 46.97 | 3.51 | 5.64 | 3.24 | 3.29 |
| | sAFM-AABB | FM | 27.36 | 38.91 | 3.47 | 5.78 | 3.20 | 3.26 |
| | sAFM-AABB | AFM | 27.58 | 39.70 | 3.49 | 5.73 | 3.21 | 3.30 |

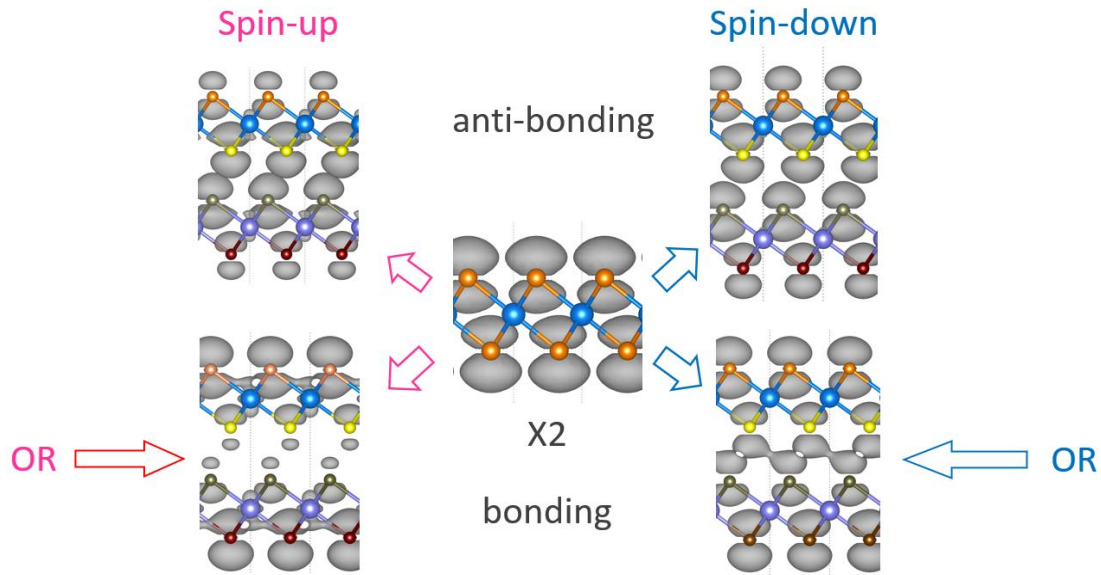

**Figure S4. Illustration and definition of the overlapped region using real-space distribution of wavefunction norms in bilayer CrSe$_2$.** Here, we show two interlayer Se $p_z$ orbitals could form a bonding (lower side of the figure) and an antibonding (upper side) orbitals at the interlayer region for both spin-up (red, left side) and –down (blue, right side) components. An isosurface value of 0.0015 e/Bohr$^3$ was used. The overlapped region is located at the area where interlayer Se $p_z$ orbitals appreciably overlapped, as marked by the red and blue horizontal arrows with notation ``OR".

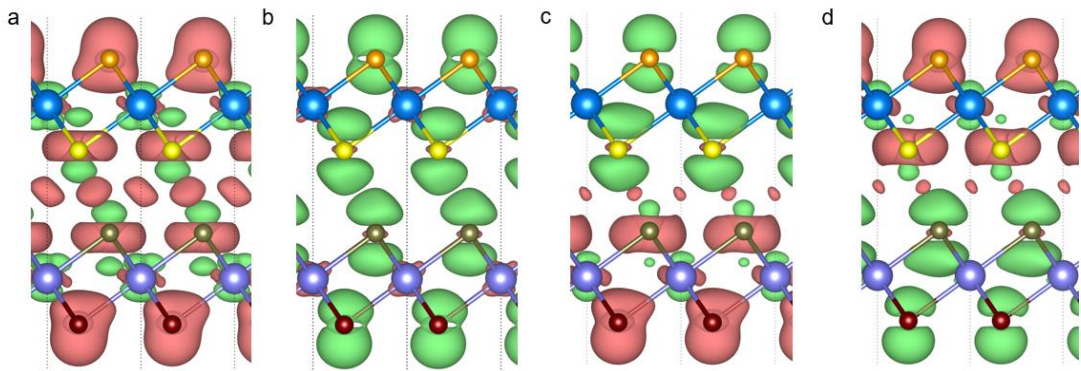

**Figure S5. Side views of interlayer differential charge densities of spin-up (a, c) and -down (b, d) of the interlayer FM (a,b) and AFM (c,d) configurations with an isosurface value of 0.001 e/Bohr$^3$.** Red and green contours indicate charge accumulation and reduction, respectively. The spin-1 density (Fig. S6a, we use spin-up for simplification hereinafter) shows charge reduction of interfacial Se $4p_z$ orbitals, accumulated charge at the OR and the Cr $e_g$ to $t_{2g}$ charge transfer. Although the spin-2 (spin-down hereinafter) density is reduced from the $4p_z$ orbitals of all Se atoms, it is still significant at the largely overlapped OR region, which is shown in the spin-density

plot (Fig. 3d), thus giving rise to spin-down polarized OR. Both plots confirm the spin-up polarized Cr (~3.11 μB) and spin-down polarized interface Se atoms (~0.2 μB) and the OR (see Fig. 3c).

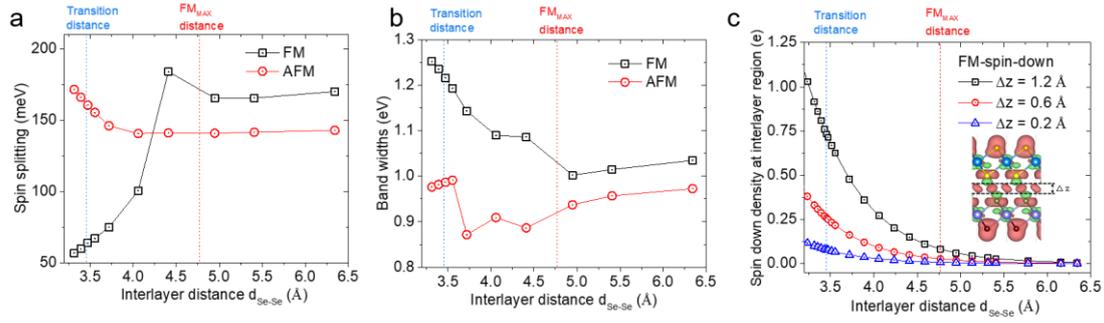

**Figure S6. Interlayer-distance dependent exchange-splitting (Coulomb), bandwidth (hopping) and electron density of spin-down (Pauli) of the interlayer bonding state.** As shown in Figure S4, Se $p_z$ orbitals overlap at the interlayer region forming the overlapped region (OR). We summarized the evolution of averaged spin splitting (a) and bandwidth (b) of the interlayer bonding state as a function of interlayer Se-Se distance $d_{Se-Se}$. At shorter interlayer distances (< 4 Å), the averaged single-orbital spin splitting (on-site U) of the interlayer bonding state in the AFM configuration are over three times that in the FM configuration. However, they rapidly approach to each other that the splitting of AFM decays but abruptly increases for FM with increasing interlayer distance, as shown in (a). A crossover occurs before they getting saturated at close values. In terms of the bandwidth (kinetic energy across layer), we choose the majority spin of Se for data counting. The values of the both configurations do not significantly change as a function of interlayer distance and the AFM values are always smaller than, ~80-90%, the FM values at certain distances. The electron density of spin-down (c) are integrated in a slab positioned at the interlayer middle plane with thicknesses $\Delta z$ of 0.2 (blue line), 0.6 (red line) and 1.2 (black line) Å. The density abruptly increases with decreasing interlayer distance, leading to rapidly enhanced Pauli repulsion and the bilayer thus favors the AFM order at shorter interlayer distances. All the splitting, bandwidth and spin-down electron density show saturating behaviors at the interlayer distance beyond 5 Å, ascribed to the suppressed interlayer wavefunction overlap in largely separated layers.

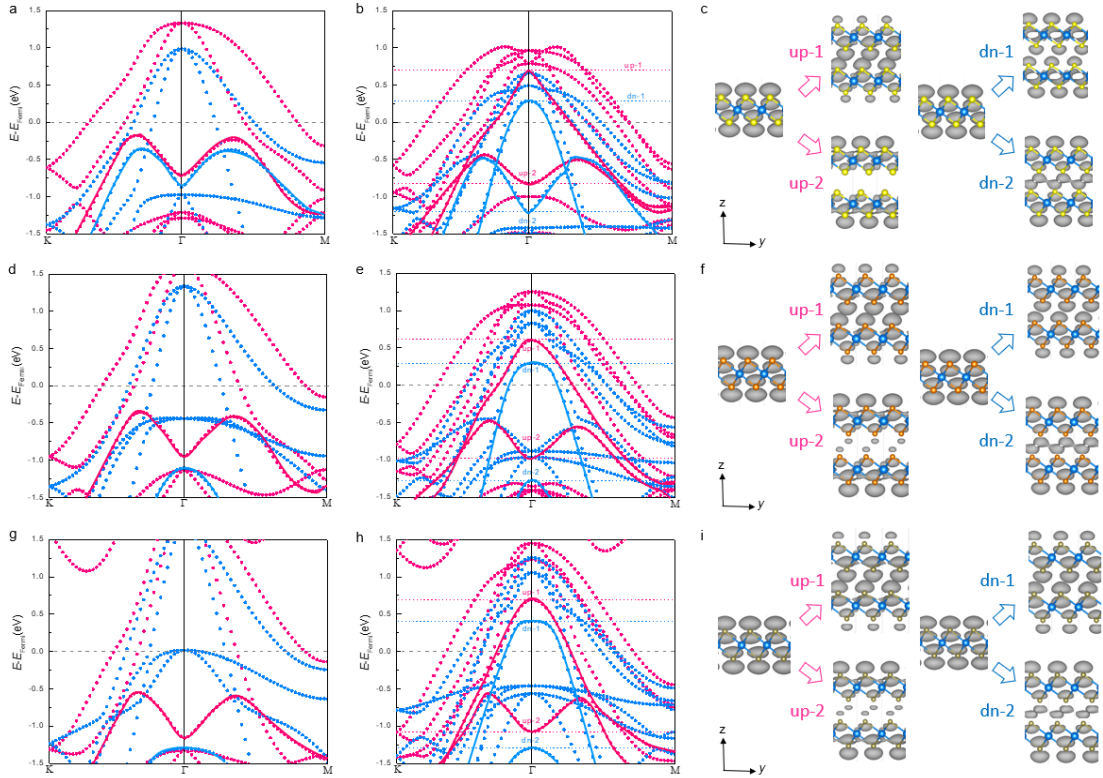

**Figure S7. Band splitting from 1L to 2L of CrX$_2$ (X=S, Se, Te).** (a-b) Spin polarized bandstructures of 1L and 2L CrS$_2$, the red and blue bands correspond to spin-up and spin-down electrons. Respectively the bands corresponding to the formation of covalence-like quasi bonds were highlighted spanning the Brillouin zone by red solid lines and blue solid lines. The Fermi level was marked using grey dashed line. The values of these two splitting were shown in **Table S3**. (d-e) were for CrSe$_2$ and (g-h) were for CrTe$_2$. (c) Visualized wave-functions for the labeled states in the 1L and 2L CrS$_2$ with an isosurface value of 0.0015 e/Bohr$^3$, while (f) for CrSe$_2$ and (i) for CrTe$_2$.

When two CrX$_2$ layers approach together under vdW attractions, there are two, *i.e.* Pauli and Coulomb, repulsive interactions to balance the vdW attractions. The Pauli repulsion is a pure quantum result of electron degeneracy, which is more significant at shorter distances where substantial wavefunctions overlap occurs, while the Coulomb repulsion works in a much wider range of distance. The strengths of these two repulsions could be evaluated quantitively by the effective interlayer crystal field splitting (CFS) and the Coulomb splitting (CS) in the bandstructures. While the CS is well defined as the energy difference of spin-up and –down components at the Γ point ($E_{up-1}$ - $E_{dn-1}$ and $E_{up-2}$ - $E_{dn-2}$), the CFS is defined as the splitting of antibonding ($E_{up-1}$ and $E_{dn-1}$) and bonding ($E_{up-2}$ and $E_{dn-2}$) states when two monolayers are stacked together, which are relevant to Pauli repulsion. Both repulsive interactions show an opposite trend to each other. Among all these CrX$_2$ bilayers, the CrS$_2$ bilayer has the strongest Coulomb repulsion (CS=0.39-0.41 eV) and weakest Pauli repulsion (CFS=1.49-1.52 eV). The Coulomb repulsion thus prevents the two CrS$_2$ layers from reaching the closer Pauli regime. However, Pauli repulsion, with increased CFS to 1.70-1.77 eV, substantially contributes to the interlayer repulsion of the CrTe$_2$ bilayer, which is evidenced by the reduced Coulomb repulsion (0.22 – 0.29 eV) and shortened interlayer distance. As a result, the $CrS_2$ bilayer strongly favors interlayer FM and the $CrTe_2$ bilayer largely prefers interlayer AFM, while in between them, $CrSe_2$ takes AFM with a rather small AFM-FM energy difference.

**Table S3.** Values of interlayer equivalent crystalline field splitting and Coulomb splitting. We defined the interlayer equivalent crystalline field splitting as the splitting of up-1 and up-2, dn-1 and dn-2. The Coulomb splitting was defined as the splitting of up-1 and dn-1, up-2 and dn-2. Further details were marked in Fig. S6. The crystal field splitting increases for both spin-up and spin-down electrons from S atom to Te, approximatively illustrating the increasing of the interlayer Pauli repulsion. In contrast, the Coulomb splitting decreases from S atom to Te atom, demonstrating the decreasing of the Coulomb repulsion.

|        | Crystal field splitting (eV) |       | Coulomb splitting (eV) |      |
| :----: | :----: | :----: | :----: | :----: |
|        | up   | down | 1    | 2    |
| $CrS_2$  | 1.52 | 1.49 | 0.41 | 0.39 |
| $CrSe_2$ | 1.58 | 1.56 | 0.31 | 0.29 |
| $CrTe_2$ | 1.77 | 1.70 | 0.29 | 0.22 |

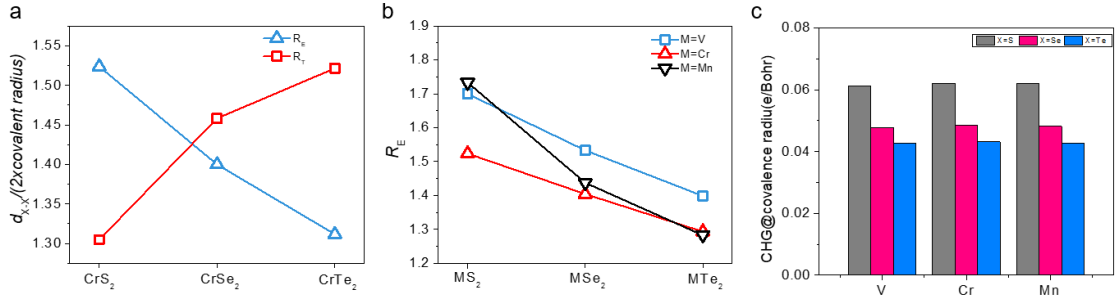

**Figure S8. The semi-quantitative description of the Coulomb repulsion of MX$_2$ (M=V, Cr, Mn; X=S, Se, Te).** (a) The ratio of the equilibrium distances (blue hollow triangles) and the transition distances (red hollow squares) of CrX$_2$ (X=S, Se, Te) compared to double the respective covalent radius. The equilibrium distances are gained from the geometric relaxation using optB86b-vdW, while the transition distance are chosen from the HSE06 results in Fig.4c. (b) The ratio of the equilibrium interlayer ($R_E$) X-X distance of MX$_2$ compared to double the respective covalent radius. The equilibrium distances are gained from the geometric relaxation using PBE-dDsC+UJ. The S atom has the smallest atomic and covalent radii as well as the largest charge density around the covalence radius, illustrating the strongest Coulomb repulsion and resulting in the largest $R_E$. (c) The summation of the electronic charge density in the plane which is as far as the length of the covalent radius away from the nonmetallic atoms of the 1L MX$_2$ (M=V, Cr, Mn; X=S, Se, Te). To a certain metal atom, the charge density decreases from S to Te. However, the type of metal atom shows little effects on the charge density.

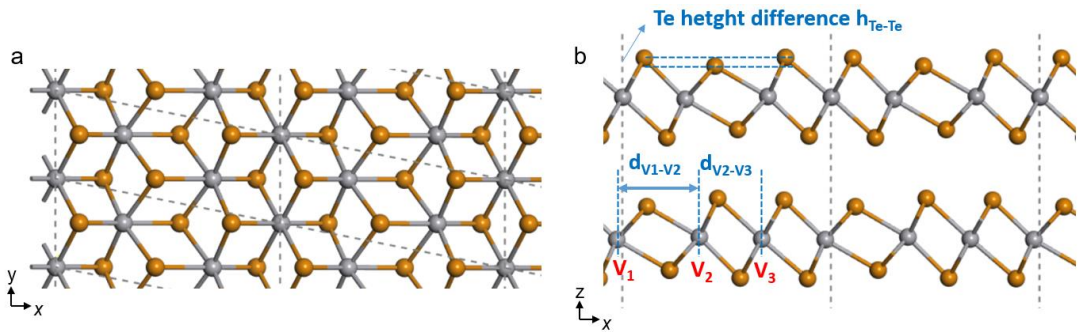

**Figure S9. Top and side view of the 1T" phase VTe$_2$.** 1T"-VTe$_2$ is energetically more stable than 1T-VTe$_2$. Characteristic geometric parameters are marked in Fig. S9b and listed in Table S4.

**Table S4.** Relative total energies, lattice constants *a* and *b*, interlayer distance, bond lengths and vertical height differences marked in Fig. S9 of bilayer VTe$_2$ in the 1T and 1T" phases. VTe$_2$ was, as a special case, found to be in a monoclinic 1T" phase, which is 35.2 meV/V more stable than its 1T phase. The BSC-like behavior maintains in this distorted phase, suggesting its robustness.

|  | ΔE (meV/V) | a (Å) | b (Å) | Interlayer distance Te-Te (Å) | $d_{V1-V2}$ (Å) | $d_{V2-V3}$ (Å) | $h_{Te-Te}$ (Å) |
|---|---|---|---|---|---|---|---|
| 1T-1x1 | +35.2 | 3.67 | 3.67 | 3.22 | 3.67 | 3.67 | 0 |
| 1T"-1x3 | 0.0 | 3.68 | 3.75 | 3.08 | 4.22 | 3.46 | 0.36 |

**Table S5.** Transition interlayer Se-Se distance of CrSe$_2$ calculated using different functionals, including PBE+UJ (+SOC), PBE-dDsC+UJ, optB86b-vdW+UJ and HSE06. All the functionals give relevant results.

| Bilayer CrSe$_2$ | Transition distance(Å) |
|---|---|
| PBE+UJ | 3.45 |
| PBE+UJ+SOC | 3.43 |
| PBE-dDsC +UJ | 3.43 |
| optB86b-vdW+UJ | 3.53 |
| HSE06 | 3.50 |

**Table S6.** Magnetic anisotropy energies (MAEs) for FM-FM and FM-AFM states of AA-stacked CrX$_2$ (X = S, Se, Te) bilayers.

| MX$_2$-2L | Mag. Config. | | Easy axis | MAE (meV/Cr) |
|---|---|---|---|---|
|  | intralayer | interlayer |  |  |
| CrS$_2$ | FM | FM | *y* | 0.006 |
|  | FM | FM | *z* | 0.040 |
| CrSe$_2$ | FM | FM | *z* | 0.139 |
|  | FM | AFM | *x* | 0.062 |
| CrTe$_2$ | FM | FM | *y* | 0.026 |
|  | FM | AFM | *x* | 0.043 |

**Table S7.** Equilibrium and transition interlayer Se-Se distance of $CrSe_2$ calculated using different on-site $U$ values, magnetic moments on Cr atoms are also included respectively. Larger $U$ indicates larger Pauli repulsion, causing larger transition distance and larger magnetic moments on Cr atoms.

| On-site $U$ (eV) | Equilibrium distance (Å) | | Transition distance(Å) | Mag. Mom. ($\mu_B$)/Cr | |
|---|---|---|---|---|---|
| | FM | AFM | | FM | AFM |
| 2.0 | 3.37 | 3.38 | 3.01 | 2.74 | 2.77 |
| 3.0 | 3.39 | 3.37 | 3.17 | 2.90 | 2.93 |
| 4.0 | 3.40 | 3.36 | 3.38 | 3.04 | 3.07 |
| 4.5 | 3.39 | 3.36 | 3.45 | 3.10 | 3.14 |
| 5.0 | 3.39 | 3.36 | -- | 3.17 | 3.20 |

**Table S8.** Transition interlayer X-X (X=S, Se, Te) distance of $CrX_2$ calculated with PBE and PBE+SOC. Keep $J = 0.6$ eV, the on-site $U$ calculated with linear response method are considered. Smaller U is also checked for $CrTe_2$ and the results of $U = 3.0$ eV shift closer to the results calculated with HSE06+SOC (4.20 Å), indicating linear response method might give large results to bilayer $CrTe_2$. The transition distance is highly relevant to on-site $U$ and hopping $t$ and could be affected by structural distortions usually found in metal tellurides [23-27]. Our results do not mean all prepared $CrTe_2$ bilayer samples should be AFM.

| Bilayer $CrX_2$ | Transition distance(Å) | |
|---|---|---|
| | PBE+UJ | PBE+UJ+SOC |
| $CrS_2$ (U=4.6 eV) | 2.80 | 2.80 |
| $CrSe_2$ (U=4.5 eV) | 3.45 | 3.43 |
| $CrTe_2$ (U=4.4 eV) | 5.06 | 4.65 |
| $CrTe_2$ (U=3.0 eV) | 4.32 | 4.22 |

**Table S9.** Relative total energy $\Delta E_0$ to the most stable configuration of each bilayer, interlayer distance $d_{X-X}$, transition distance $d_T$, lattice constant $a$, binding energy and magnetic moment, lattice constants $a$ and $b$ of bilayer MX$_2$ in the 1T phase. Term IMC stands for Interlayer Magnetic Coupling. Here, the AA stacking is found the most stable in all MX$_2$ bilayers, except for the VTe$_2$ bilayer as it was proved to be more stable in the FM 1T''phase.

| MX2 | Stacking | IMC. | $\Delta E_0$ (meV/M) | $d_{X-X}$ (Å) | $d_T$ (Å) | $a$ (Å) | $E_b$ (meV/Å$^2$) | Mag. Mom. (μB/M) |
|---|---|---|---|---|---|---|---|---|
| VS$_2$ | AA | **FM** | **0.00** | **3.57** | 2.80 | **3.31** | **-16.68** | 1.21 |
| | | AFM | 0.99 | 3.60 | | 3.31 | -16.48 | 1.21 |
| | AB | FM | 6.18 | 3.67 | | 3.31 | -15.38 | 1.21 |
| | | AFM | 5.73 | 3.77 | | 3.31 | -15.47 | 1.21 |
| VSe$_2$ | AA | **FM** | **0.00** | **3.68** | 3.09 | **3.45** | **-16.36** | 1.35 |
| | | AFM | 4.60 | 3.61 | | 3.45 | -15.44 | 1.34 |
| | AB | FM | 5.00 | 3.79 | | 3.44 | -15.44 | 1.36 |
| | | AFM | 6.67 | 3.84 | | 3.44 | -15.09 | 1.34 |
| CrS$_2$ | AA | **FM** | **0.00** | **3.20** | 2.78 | **3.38** | **-28.63** | 2.96 |
| | | AFM | 5.24 | 3.19 | | 3.39 | -27.39 | 2.99 |
| | AB | FM | 35.19 | 3.38 | | 3.36 | -21.72 | 2.96 |
| | | AFM | 37.86 | 3.38 | | 3.37 | -21.07 | 2.99 |
| CrSe$_2$ | AA | FM | 2.17 | 3.39 | 3.42 | 3.53 | -19.37 | 3.10 |
| | | **AFM** | **0.00** | **3.37** | | **3.54** | **-19.77** | 3.14 |
| | AB | FM | 25.99 | 3.59 | | 3.49 | -15.32 | 3.10 |
| | | AFM | 24.87 | 3.58 | | 3.49 | -15.54 | 3.12 |
| CrTe$_2$ | AA | FM | 19.85 | 3.66 | 4.58 | 3.81 | -17.86 | 3.31 |
| | | **AFM** | **0.00** | **3.57** | | **3.84** | **-20.71** | 3.38 |
| | AB | FM | 43.29 | 3.82 | | 3.77 | -14.48 | 3.30 |
| | | AFM | 32.20 | 3.76 | | 3.78 | -16.14 | 3.36 |
| MnS$_2$ | AA | **FM** | **0.00** | **3.64** | 2.67 | **3.38** | **-15.46** | 3.31 |
| | | AFM | 0.59 | 3.66 | | 3.38 | -15.34 | 3.31 |
| | AB | FM | 2.34 | 3.73 | | 3.38 | -15.01 | 3.30 |
| | | AFM | 0.22 | 3.70 | | 3.38 | -15.39 | 3.31 |
| MnSe$_2$ | AA | **FM** | **0.00** | **3.45** | 2.80 | **3.59** | **-18.98** | 3.73 |
| | | AFM | 25.30 | 3.63 | | 3.60 | -14.36 | 3.67 |
| | AB | FM | 13.32 | 3.63 | | 3.57 | -16.73 | 3.69 |
| | | AFM | 20.84 | 3.77 | | 3.58 | -15.32 | 3.66 |
| MnTe$_2$ | AA | **FM** | **0.00** | **3.54** | 2.94 | **3.87** | **-20.42** | 3.73 |
| | | AFM | 47.04 | 3.50 | | 3.95 | -12.63 | 3.75 |
| | AB | FM | 18.28 | 3.79 | | 3.81 | -18.17 | 3.68 |
| | | AFM | 28.97 | 3.81 | | 3.80 | -16.52 | 3.68 |

**Table S10.** Geometric structure and magnetic of the bilayer $CrSe_2$ calculated with optB86b-vdW and PBE-dDsC, On-site $U$ and $J$ are also considered during the relaxation. Interlayer AFM to interlayer FM energy different $\Delta E$, binding energy of the bilayer $CrSe_2$, intralayer lattice constants $a$ for hexangular unit cell, the thickness of layer $\Delta c$ and the perpendicular distance between the interlayer Se atoms $\Delta d$, magnetic moments of different atoms are included. These two functional almost give the same results. Top view, side view and legends are provided below. The $\Delta E$ here are calculated with the corresponding functional, indicating interlayer AFM is energetically more stable. We choose the $\Delta E$ calculated with PBE+UJ with the structure of optB86b-vdW+UJ in the main text. To evaluate how the two functional affect the magnetic moment, we choose the moments calculated by corresponding functional, which is different to Table S1.

| Functioanl | Intra/Interlayer Mag. Config. | $\Delta E$ (meV/Cr) | Str. constants (Å) | | | Mag. Mom. (µB) | | |
|---|---|---|---|---|---|---|---|---|
| | | | a | $\Delta c$ | $\Delta d$ | Cr | Se_ot | Se_it |
| optB86b-vdW | FM/FM | 1.85 | 3.53 | 3.01 | 2.73 | 3.11 | -0.20 | -0.22 |
| | FM/AFM | 0.00 | 3.54 | 2.99 | 2.67 | 3.14 | -0.21 | -0.20 |
| PBE-dDsC | FM/FM | 2.17 | 3.53 | 3.00 | 2.71 | 3.10 | -0.20 | -0.22 |
| | FM/AFM | 0.00 | 3.54 | 3.00 | 2.67 | 3.14 | -0.21 | -0.20 |

**Table S11.** On site Coulomb $U$ and exchange $J$ of monolayer $VX_2$, $CrX_2$, $MnX_2$ and $CrI_3$. These values were calculated with a linear response method based on the magnetic ground state of these monolayers. We used the effective $U$ in $VX_2$ given the small magnetism moments resulting in large errors of $J$ values.

| $MX_2$ | $U$ (eV) | $J$ (eV) |
|---|---|---|
| $VS_2$ | 3.1 | -- |
| $VSe_2$ | 2.7 | -- |
| $VTe_2$ | 2.2 | -- |
| $CrS_2$ | 4.6 | 0.6 |
| $CrSe_2$ | 4.5 | 0.6 |
| $CrTe_2$ | 4.4 | 0.6 |
| $MnS_2$ | 4.1 | 0.8 |
| $MnSe_2$ | 4.0 | 0.7 |
| $MnTe_2$ | 3.0 | 0.7 |
| $CrI_3$ | 3.9 | 1.1 |